\documentclass{article}
\def\be{\begin{equation}}
\def\ee{\end{equation}}
\def\bea{\begin{eqnarray}}
\def\eea{\end{eqnarray}}

\def\4pig{\sfrac{4\pi G}{c^{4}}}

\def\hsp5{\hspace{5mm}}
\newcommand{\sfrac}[2]{{\textstyle{#1\over#2}}}
\def\case#1/#2{\textstyle\frac{#1}{#2}}

\begin{document}

\title{Inhomogeneity effects in Cosmology}
\author{George F R Ellis\footnote{E-mail: george.ellis@uct.ac.za} \\
\\
ACGC and Department of Mathematics and Applied Mathematics,\\
University of Cape Town, \\Rondebosch 7701, Cape Town, South
Africa.}

\maketitle


\begin{abstract}
This article looks  at how inhomogeneous spacetime models may be
significant for cosmology.  First it looks at how the averaging
process may affect large scale dynamics, with backreaction effects
leading to effective contributions to the averaged energy-momentum
tensor. Secondly it considers how local inhomogeneities may affect
cosmological observations in cosmology, possibly significantly
affecting the concordance model parameters. Thirdly it presents the
possibility that the universe is spatially inhomogeneous on Hubble
scales, with a violation of the Copernican principle leading to an
apparent acceleration of the universe. This could  perhaps even
remove the need for the postulate of dark energy.
\end{abstract}

\section{Introduction}
The standard models of present day cosmology are perturbed
FLRW (Friedmann-Lema\^{\i}tre-Robertson-Walker) models. These
models, developed by Einstein, de Sitter, Friedmann,
Lema\^{\i}tre, Robertson, and Walker in the period from 1917 to
1935, are exactly spatially homogeneous and isotropic, with an
implied smooth fluid approximation; an early standard reference on
their properties is by Robertson \cite{Rob33}. The main
developments since then are, firstly, consideration of much more
complex matter content than considered at that time, in particular
considering inclusion of background radiation interacting with
multiple matter components and scalar fields, allowing in particular
an inflationary early epoch; secondly, and consequent on this, a
sophisticated history of the physical evolution of the contents of
the universe, including in particular nucleosynthesis and
matter-radiation decoupling; and thirdly, following the pioneering
work of Lifschitz, the extension of these models to perturbed models
where linearized structure formation and its effects on the
background radiation can be studied. Observational relations can be
calculated in these models and compared with astronomical data,
confirming that they give good physical models that account
satisfactorily for these observations. Summaries are given in many
texts, e.g. Dodelson \cite{Dod03}, Peters and Uzan 
\cite{PetUza09}, Ellis, Maartens and MacCallum 
\cite{EllMaaMac11}.

The basic model is very successful, but has major mysteries:
particular the nature of dark matter on the one hand, and the nature
of the dark energy causing acceleration of the universe at recent
times on the other. However like all models, it is an idealization:
it represents the background model and linear perturbations around
it very well, but the real universe has non-linear structure and
voids at scales smaller than the Hubble scale \cite{web}, which
are not well represented by these models.

The FLRW model is a large-scale approximation to these non-linear
structures, that is supposed to represent the result of global averaging of
inhomogeneities. There are three key issues here:
\begin{itemize}
  \item local inhomogeneities may affect the averaged large
scale dynamics,
  \item locally inhomogeneities affect photon propagation, and so may affect cosmological observations,
  \item maybe the universe is after all not spatially homogeneous on the
  largest scales, and is better represented at late times by a Lema\^{\i}tre-Tolman-Bondi (LTB)
  spherically symmetric model, where we are situated near the centre of a Hubble scale void.
\end{itemize}
These concerns, which are not mutually exclusive, gain traction
because of the mysterious issue of dark energy, whose nature is
completely unknown. So the question is not just whether
inhomogeneities may significantly affect the interpretation of
observations in cosmology; it is whether they can affect the need
for dark energy, or at least significantly affect the concordance
model parameter values. In brief: is inhomogeneity important for
cosmology itself, apart from being central to the study of structure
growth?

These are the issues I shall introduce here. There is a large
literature on these topics, so I can only refer to representative
publications on them in the following sections; most of the
relevant papers will be mentioned in the further articles in this
special issue. Note that this is not an article on the use of
inhomogeneous models to explore structure formation in the expanding
universe: that is a separate, though related, issue.

\subsection{Preliminaries}
The Einstein Field Equations (`EFE') algebraically determine
$R_{ab}$ from the matter tensor $T_{ab}$: \footnote{ Geometrized
units, characterized by $c = 1 = 8\pi G/c^{2}$, are used
throughout.}
\begin{equation}  \label{efe}
R_{ab} = T_{ab} - {\textstyle{\frac{1}{2}}}\,T\,g_{ab} +
\Lambda\,g_{ab} \hsp 5 \Rightarrow \hsp5 R = -\,T + 4\,\Lambda \ .
\end{equation}
When the matter takes a `perfect fluid' form, 
\begin{equation}\label{pf}
T_{ab} = (\mu+p)\,u_{a}\,u_{b} + p\,g_{ab} \hsp5 \Rightarrow \hsp5 T
= -\,(\mu-3p)
\end{equation}
with $\mu$ the total energy density and $p$ the isotropic pressure,
the Ricci tensor expression is
\begin{equation} \label{ricci}
R_{ab} = (\mu+p)\,u_{a}\,u_{b} +
{\textstyle{\frac{1}{2}}}\,(\mu-p+2\Lambda) \,g_{ab} \hsp5
\Rightarrow \hsp5 R = (\mu-3p) + 4\,\Lambda \ .
\end{equation}
This is necessarily the case in a FLRW model.
The cosmological constant $\Lambda$ is equivalent to a Ricci tensor
contribution (\ref{ricci})  with $\mu_\Lambda+p_\Lambda=0$. That is
one can represent $\Lambda$ either on the left hand side of the EFE
as in (\ref{efe}), or on the right hand side of the EFE  as a fluid
(\ref{pf}) with equation of state $p = - \mu$.

\section{Backreaction effects}
\label{sec:back}

\subsection{The basic idea}
The concept of backreaction from smaller to larger scales was
developed in a paper by Brill and Hartle \cite{BriHar64} in the
context of John Wheeler's idea of geons. It was
extended to the case of gravitational radiation in two beautiful
papers by Isaacson \cite{Isa68a,Isa68b}. He envisaged high frequency
waves superimposed on a slowly varying background:
\begin{equation}\label{isaacson1}
g_{\mu\nu} = \gamma_{\mu\nu} + \epsilon h_{\mu\nu}, \,\,\partial
\gamma \simeq \gamma/L, \,\,\,\partial h \simeq h/\lambda, \,
\end{equation}
where $L$ is the lengthscale of variation of the background metric
and $\lambda$ that of the gravitational waves superimposed on the
background. Then
\begin{equation}\label{isaacson2}
R_{\mu\nu}(\gamma + \epsilon h) = R_{\mu\nu}^{(0)} + \epsilon
R_{\mu\nu}^{(1)} + \epsilon^{2} R_{\mu\nu}^{(2)} + \epsilon^3
R_{\mu\nu}^{(3+)}
\end{equation}
where $R_{\mu\nu}^{(0)} = R_{\mu\nu}^{(0)}(\gamma)$ and the others
are functions of $h_{\mu\nu}$. Thus if the actual spacetime is
empty: $R_{\mu\nu}(g) = 0$, the background metric is not that of an
empty spacetime: $R_{\mu\nu}^{(0)} \neq 0$ and there is an effective matter term on the right hand
side of the EFE. One finds $R_{\mu\nu}^{(1)}=0$ and
\begin{eqnarray}\label{isaacson3}
R_{\mu\nu}^{(0)} - \frac{1}{2}R^{(0)} \gamma_{\mu\nu} = -8\pi T_{\mu\nu}^{\,\,\,eff},\\
T_{\mu\nu}^{\,\,\,eff} = \frac{\epsilon^2}{8\pi}( R_{\mu\nu}^{(2)}
-\frac{1}{2} R^{(2)} \gamma_{\mu\nu} )
\end{eqnarray} so the
gravitational wave appears as a source of the background. This is
backreaction from the small scale structure to the large scale
structure.

This illustrates the basic backreaction proposal: coarse-graining
microstructure results in effective matter components at macro
scales that can influence the macro (coarse-grained) dynamics. The
issue was taken up {\it inter alia} by Szekeres \cite{Sze71}, who showed
that this averaging effect could be expressed in a weak-field
polarization formalism in analogy with the electromagnetic case; by
MacCallum and Taub \cite{MacTau73}, who derived Isaacson's results
using a two-time Lagrangian formalism; and by Noonan \cite{Noo84},
who extended Isaacson's formulation to include matter (an
astronomical ``medium'').

\subsection{Non-commutativity of EFE and averaging}

The basic point is that averaging the geometry and calculating the
field equations do not commute \cite{Ell84,EllBuc05}. We use angle
brackets to denote averaging over a suitable volume $V$, so
$\overline{g}_{ab} \equiv \langle g_{ab}\rangle$ is the background
metric with inverse $\overline{g}^{ab}$ given by $\overline{g}^{ab}
\overline{g}_{bc}=\delta^a_c$, and indices should be raised and
lowered using the full metric $g^{ab}$, $g_{ab}$. Then
\begin{equation}  \label{average1}
 g_{ab} = \overline{g}_{ab} + \delta g_{ab}\,\,\,g^{ab} g_{bc} = \delta^a_c ,\,\,\,
g^{ab} = \overline{g}^{ab}  + h^{ab}
\end{equation}
shows that
\begin{equation}  \label{average4}
(\overline{g}^{ab}  + h^{ab}) (\overline{g}_{bc} + \delta
g_{bc})=\delta^a_c
\end{equation}
so $\overline{g}^{ab} \neq \langle g^{ab}\rangle$ and ${h}^{ab} \neq
\delta g^{ab} \equiv  g^{ae} g^{bf} (\delta g_{ef})$. Consequently the
Christoffel symbols gain extra terms relative to the averaged Christoffel symbols: $\Gamma^a_{\,\,\,bc}
= \overline{\Gamma}^a_{\,\,\,bc} + \delta \Gamma^a_{\,\,\,bc}$ and
the Ricci and Einstein tensors in turn gain extra terms:
\begin{equation}  \label{average2}
R_{ab} = \overline{R}_{ab} + \delta R_{ab}\,\,
 \Rightarrow  G_{ab} = \overline{G}_{ab}  + \delta G_{ab}\,,
\end{equation}
hence the averaged EFE gain an extra term:
\begin{equation}  \label{average3}
G_{ab} = T_{ab}  + \Lambda\,g_{ab} \Rightarrow \overline{G}_{ab} +
\delta G_{ab} =
\overline{T}_{ab}  + \Lambda\,g_{ab} 
\end{equation}
These extra terms are effective matter terms in the large scale
field equations, consequent on the coarse-graining of small scale
inhomogeneities; this called the backreaction from the smaller to
the larger scales, and is consequent on the fact that
coarse-graining (or averaging) does not commute with calculating the
EFE from the metric tensor:
\begin{equation}  \label{average5}
\overline{G}_{ab} = \widetilde{T}_{ab}  + \Lambda\,g_{ab},\,\,
\widetilde{T}_{ab} := \overline{T}_{ab}  
- \delta G_{ab}
\end{equation}
where $\widetilde{T}_{ab}$ is the effective coarse-grained source
term, the second term on the right being the effect of matter
averaging and the third term the geometric backreaction effect. The
Isaacson gravitational radiation calculation summarised above is a
specific example (a vacuum spacetime with a rapidly varying
gravitational wave appears to have an effective matter content when
viewed on larger scales).

In principle carrying out that calculation is straightforward but
very complex. However to be certain of the result, one needs to
average in the real universe, not the background spacetime. The
basic problem then is that averaging involves integration of tensor
quantities over a spacetime volume, and so is not a well-defined
tensorial operation: changing the coordinates will change the result
in an arbitrary way. One can try to handle this by
\begin{enumerate}
  \item defining a covariant averaging of tensors via bitensors, or
  \item using only field equations involving averaged scalars, perhaps
  involving a convolution rather than simple averaging, or
  \item carrying out the calculation in a weak field approximation where the
integrals can be well defined in a highly symmetric background
spacetime, and the difference between the integral in the background
spacetime and the real spacetime is negligible, or
\item choosing a uniquely defined physically motivated coordinate
system in the fully non-linear spacetime.
  \end{enumerate}
All have been tried. There are problems in each case:
\begin{enumerate}
  \item there is no uniquely defined usable bitensor, as the Synge
  parallel transport bitensor does not work (it leaves the metric
  tensor invariant);
  \item it is not easy to find well-defined scalars that fully define the geometry and dynamics
  in a generic case;
  \item the linearized procedure may not accurately reflect the needed integral in the real
  spacetime;
  \item one is breaking general covariance in this procedure; one has to motivate
  that the result is physically meaningful.
\end{enumerate}

\subsection{Cosmological applications: Fitting and averaging}
The application of the idea of backreaction to cosmology was raised
in \cite{Ell84}, see also \cite{EllSto87}, and then taken up {\it inter
alia} by Futamase \cite{Fut88, BilFut91, Fut96}, Stoeger et al \cite{StoZot92,StoEtal07}, and particularly by
Buchert and collaborators, first in the Newtonian case
\cite{BucEhl97} and then in the GR case \cite{Buc99}. The
implications for cosmology have been discussed more recently by
Kolb, Matarrese, Wiltshire, R\"{a}s\"{a}nen, Sussman, and others.

The key point from the discussion above is that backreaction from
small scale inhomogeneity to the large scale geometry can generate a
dynamic effect in the effective Friedmann equation for the
cosmology, allowing an acceleration contribution due to back reaction from "small
scale" inhomogeneities. This has a potential effect on cosmological
parameters \cite{Bucetal00,Buc08}; the question is whether it is
large enough to give a significant contribution to dark energy. Kolb
and Wiltshire propose it can provide a sufficient source of all the
effective dark energy, leading to the possibility of concordance
cosmology without $\Lambda$. In contrast, as discussed below, many
others deny the effect is important.

 The further issue that arises is that while some form of averaging process is in
principle what one should do to arrive at the large scale geometry
of the universe on the basis of observations, in practice what is
normally done is the inverse. One assumes \emph{a priori} a FLRW
model as a background model, and then uses some form of
observationally-based fitting process to determine its basic
parameters  \cite{EllSto87}. This in effect defines a mapping from the smooth background
model into the perturbed more realistic space time,
which then \emph{defines} the specific perturbations that occur
about the background model \cite{EllBru89}, for if you change the fitting - what is
often called the \emph{gauge} - you change the perturbations.

Now there are many ways one can conceive of to perform such a
fitting, and indeed averaging is one of them: in principle one can
average energy densities, pressures, expansion rates, etc to arrive
at a FLRW model from a more accurate representation. However in
practice fitting is done via astronomical observations down the past
null cone, leading to fitting procedures for the FLRW parameters as
set out in the paradigmatic paper by Sandage \cite{San61}, updated
by all the myriad other data now used to determine the best-fit FLRW
model \cite{triangle}. Once one has fitted a specific FLRW model to
the observable region of the universe, one can then try to determine
the specific local deviations from the background model - as for
example in all the studies trying to identify the Great Attractor
\cite{GA1,GA2,GA3}. Ideally what one would do is show that both a
coarse-graining procedure and a suitable fitting procedure for a
realistic lumpy universe model - depicting all the great walls,
voids, etc - would give the same result. No one has so far shown how
this might work.

An interesting question here is whether (i) there is a scale above
which the universe is exactly FLRW, or (ii) at all scales the
universe is only ever approximately FLRW. In fact while averaging
can in principle lead to an almost homogeneous model to any degree of
approximation, it can never lead to exact homogeneity, if the
initial model is not homogeneous \cite{StoEtal07}: there will always
remain residual traces of those inhomogeneities. Fitting of course
starts off with such an exactly homogeneous model. Thus in this
sense the two cannot be exact inverses of each other, and there
cannot be any scale where the universe is exactly FLRW - but it can
be very closely so.

Whether these effects are sufficient to significantly alter the
cosmological parameters determined from supernova observations
\cite{GooLei11} is an important ongoing debate involving
interesting modeling and general relativity issues, and
particularly how one models a universe with genuinely large--scale
voids, as well as the nature of the Newtonian limit in cosmology
(see the article by R\"{a}s\"{a}nen). In this section we consider various
approaches to averaging and determining backreaction.

\subsubsection{The Zalaletdinov approach} The problem with employing a
tensorial averaging procedure is that the result is not covariant:
one obtains coordinate dependent results, unless one uses bitensors
to define covariant averaging in a local domain, as proposed by
Zalaletdinov \cite{Zal92,Zal08}. This can be done for curvature and
matter, but is difficult to do in a unique way for metric itself,
because the metric is invariant under parallel propagation, so the
Synge bitensor will not work. In any case this approach leads to
complex equations  that have not yet been productive in terms of the
cosmological backreaction problem, despite some valiant attempts
\cite{ColPelZal05}.

\subsubsection{The Buchert approach}
Alternatively, one can avoid this problem by only averaging scalars,
as Buchert \cite{Buc99,Buc08} does. He shows this can in principle
provide an effective acceleration term in the averaged equations.

The key point is that expansion and averaging do not commute: in any
domain $D$, for any field $\psi$,
\begin{equation}
 \partial_t \left<\psi\right>_D - \left< \partial_t \psi\right>_D = \left<\Theta\psi\right>_D
 - \left<\theta\right>_D \left<\psi\right>_D
\end{equation}
where $\Theta$ is the expansion rate. This leads to Buchert's
modified Friedmann and Raychaudhuri equations: e.g.
\begin{equation}
\partial_t\left<\Theta\right>_D = \Lambda - 4\pi G \rho_D + 2 \left<II\right>_D - \left<I\right>_D^2
\end{equation}
where $II = \Theta^2/3 - \sigma^2 $ and $I = \Theta$, $\sigma$ being
the shear. This in principle allows acceleration terms to arise from
the averaging process, through the term $\left<II\right>_D$. To complete the dynamical equations, one
needs the shear evolution, but this cannot easily be obtained from
the full set of 1+3 dynamic equations through such averaging of
scalars. Hence Buchert's analysis relies on an Ansatz for this
evolution, which is not fully justified from the underlying
dynamics. There are integrability conditions linking the
shear to the curvature that give a combined conservation law for
curvature plus fluctuations. This forms the basis for the assumed closure
conditions, leading to exact classes of solutions where the evolution
of the averaged shear is determined. The closure condition
replaces what in Friedmannian cosmology would be the equation of
state for the sources; here it is the equation of state for the
effective sources.

The Buchert equations indicate the broad nature of the effect, and
are widely used as the basis of further studies, for instance by
Kolb et al, Wiltshire, and R\"{a}s\"{a}nen. Buchert presents his approach in
his article in this special issue. The use of scalars more generally is proposed by Coley \cite{Col10}.

\subsubsection{The renormalization group approach}

Carfora and Piotrkowska have developed a sophisticated geodesic-ball
based averaging approach, {\it inter alia} using the ideas of the
renormalization group \cite{CarPio95}. This has lead to intriguing
analyses of the effects of such averaging on cosmology
\cite{BucCar01,BucCar02,BucCar03,BucCar08}, giving formula for
averaged effects on cosmic parameters. This is a very sophisticated
extension of the basic Buchert programme, indeed it is something of
a technical \emph{tour de force} \cite{Car10}. Its relation to practical
cosmological observations is still to be developed.

\subsection{Non-linear models}
The previous approaches are not tied in to specific geometric models
of the universe. The key issue however is how good the linear models
are at representing the non-linear inhomogeneities in the real
universe, with gigantic voids, walls, and so on at larger scales,
and mainly empty space at smaller scales \cite{voids_2010}:

\begin{quotation}
Voids have been known as a feature of the Megaparsec universe since
the first galaxy redshift surveys were compiled. Voids are enormous
regions with sizes in the range of $20-50h ^{-1}$ Mpc that are
practically devoid of any galaxy, usually roundish in shape and
occupying the major share of space in the Universe [37]. Forming an
essential ingredient of the Cosmic Web, they are surrounded by
elongated filaments, sheetlike walls, and dense compact clusters.
\end{quotation}

Various non-linear models have been developed that try to
approximate this kind of situation without using a linearization
procedure; they are discussed in the articles by Bolejko, C\'{e}l\'{e}rier,
and Krasinski (and see \cite{Kra97,BKHC2009,Hel09} for discussions
of exact inhomogeneous models). The original such models were the
``Swiss cheese'' models of Einstein and Straus
\cite{EinStr45,EinStr45a}, where spherical `vacuoles' with a
spherical mass at the centre are cut out of an expanding FLRW
universe model. This gives an exact solution of the EFE with static
voids imbedded in an expanding universe. However there is no
dynamical backreaction from the inhomogeneities in these models,
because the matching conditions between the voids and the expanding
universe require that the mass at the centre of each vacuole is the
same as would have been there if there were no vacuole.

Models with voids have been developed in depth by Wiltshire
\cite{Wil07,Wil08,Wil09} who has emphasized that time runs
differently in the voids, potentially leading to a substantial
effect when integrated over long times. Furthermore voids expand
while clusters collapse or stay the same size, so the universe
becomes void dominated, and the region we live in is increasingly
not representative or "average". These models can potentially lead
to apparent acceleration of the universe \cite{SmaWil10}. However
the degree to which the models represent the real universe is not
clear. These models are discussed in the article by Wiltshire.

A completely different approach is to construct the expanding model
from an aggregation of local spherical vacuum regions, joined
together at boundary surfaces, as developed first by Wheeler and
Lindquist \cite{LinWhe57}. These models are radically different from
all the others in that here one does not start with a FLRW model and
then perturb it or excise regions from it: rather a FLRW-like
structure emerges at large scales as an approximation to the small
scale vacuum domains with imbedded static masses. Thus there is no
back-reaction to a large scale model because there was no such model
to begin with. Rather the junctions between local inhomogeneities
underlie the large scale dynamics, which is emergent rather than the
result of averaging. This approach has been developed interestingly
at recent times by Clifton and Fereira \cite{CliFer10,Cli10}. These
models are discussed in the article by Clifton (and see also \cite{UzaEllLar10}, discussed further below).

\subsection{Perturbative approach}

In contrast to these attempts at non-linear models, there is a large
literature studying backreaction effects on the basis of linearly
perturbed FLRW models. Differing views are held as to the result,
reviewed recently by Clarkson and Maartens \cite{ClaMaa10} (and see also \cite{Claetal11}). Some
workers claim the weak field approximation is adequate to describe
the non-linear structures, because the gravitational potential is
very small even though the density contrasts are very large; and
consequently the backreaction effect is negligible (see
\cite{Bauetal10} for this view). Counter claim by Kolb, Wiltshire,
Matarrese and others (see e.g. \cite{KolMatRio05,Wil11}) emphasize that as there are major voids in the
expanding universe, a weak-field kind of approximation to a
spatially homogeneous model is not adequate: you have to properly
model (possibly quasi-static) voids and their junctions to the
expanding external universe, and the linear models are not adequate
for this purpose. An in between view is given by Clarkson, Ananda,
and Larena (\cite{ClaAnaLar09}).

A recent contribution from the skeptical side is by Green and Wald
\cite{GreWal11}, using an ultra-local averaging procedure to show
- in direct contradiction of Buchert's claim - that no
acceleration can result from backreaction associated with such
averaging, because the effect is trace-free. The limiting process
embodied in this elegant work probably does not adequately
represent the results of averaging over finite volumes, as
represented by the other methods discussed here, because it does
not in fact involve any such averaging, so this method does not
disprove Buchert's results. Indeed it is unlikely that this ``trace-free'' result is true for
models that genuinely represent averaging over finite volumes.

There are many workers skeptical there is any significant effect,
with strong arguments based on the perturbed FLRW approach: the
gravitational potentials involved are so small that a
quasi-Newtonian analysis is adequate, and the back reaction effect
does indeed occur but is negligible. However others suggest it may
be at least large enough to affect the cosmic relation between
energy densities and expansion that leads us to deduce the spatial
curvature is almost flat. Greater conceptual clarity on the modeling
issues involved is required; the issue is discussed in the articles
by Kolb and Clarkson. Three specific issues arise that suggest caution
is advisable before accepting the pessimistic view.

\subsubsection{The averaging process}
In the weak field case, the perturbed quantities can be averaged
in the background unperturbed Robertson--Walker geometry: a
linearized calculations in the background spacetime. This is
central to the weak field approach. But that procedure is
inadequate for truly non--linear cases, where the integral needs
to be done over a generic lumpy (non--linearly perturbed)
spacetime that are not ``perturbations'' of a high--symmetry
background. It is precisely in these cases that the most
interesting effects will occur.

\subsubsection{Global coordinates: Models with genuine voids}
The response often given is that even though the density may be
highly non-linear, in a suitable non-comoving quasi-newtonian frame
the gravitational potential remains very small. Then one has
$\delta\rho/\rho \simeq 10^{+28}$ but $\delta \phi/\phi \simeq
10^{-5}$. This is possible because the second derivatives of the
potential are not small, and they are what enter the field equations
to balance the very large density perturbations \cite{Bucetal09}; so
a suitable linearised approach is acceptable.

Underlying this is the issue of global existence of the
quasi-Newtonian coordinates in situations of real inhomogeneity with
locally static almost empty spacetimes joining together to form an
expanding universe, as envisaged by Lindquist and Wheeler. The case
for global validity of these coordinates is put for example by
Ishibashi and Wald \cite{IshWal06} and by Baumann et al
\cite{Bauetal10}. In the Poisson gauge to second-order in scalar
perturbations the metric is
\begin{equation}\label{weakmetric}
ds^2=-\left(1+2\Phi+\Phi^{(2)}\right)dt^2 +a^2(t)
\left(1-2\Psi-\Psi^{(2)}\right)\delta_{ij}d x^i dx^j.
\end{equation}
 The first-order scalar perturbations are given by $\Phi, \Psi$, and the second-order ones by
$\Phi^{(2)}, \Psi^{(2)}$ (which are needed for a consistent analysis
of backreaction).

But the fact that such coordinates can on the one hand be used
globally in an asymptotically flat inhomogeneous region such as the
solar system, and on the other in a linearly perturbed FLRW model,
does not mean it can be used globally for a genuinely inhomogeneous
expanding universe model including both kinds of domains, as claimed
by Wald and Ishibashi. For example Lindquist and Wheeler
\cite{LinWhe57} do not give a global coordinate system: they match
local coordinates to each other across a boundary. But this is not
done exactly, because the geometry is too complex to do so. The one
case where one can do the job exactly is an expanding two-mass
solution with locally static voids joined to create an expanding
universe with compact space sections \cite{UzaEllLar10}. The
surprising result is that the join can only be done across a null
surface (a `horizon'), with intermediate spatially homogeneous
anisotropically expanding vacuum regions - it is the existence of
these regions that allow the universe to expand. It is not possible
to find global coordinates of the form (\ref{weakmetric}) in such a
spacetime, as posited by Wald and Ishibashi. Thus in that case the
weak field arguments do not apply because the coordinate system on
which they rely does not exist globally. They may however be
possible in Swiss Cheese models, where it is the intervening fluid
domains that allow the static vacuum domains to move way from each
other; but these are not the kind of situation we consider here,
with galaxies everywhere imbedded in genuinely vacuum regions and no
fluid-filled domains acting as buffers between them.

So real inhomogeneities have properties that are not the same as
perturbed FLRW models that are fluid-filled everywhere. The key
issue underlying the two-mass result is the rigidity of local
spherical vacuum regions that is embodied in Birkhoff's Theorem. So
a criticism might be that Birkhoff's theorem applies only to exact
spherically symmetry vacuum solutions; the argument won't apply to
more realistic solutions with almost spherically symmetric vacuum
domains. However this argument is invalid: an `Almost Birkhoff'
theorem shows the Birkhoff result is stable \cite{GosEll11}. On this
view, the issue is whether (on appropriate averaging scales) the
real universe is globally filled with an intergalactic medium that
can serve as the substratum allowing expansion to take place in a
way compatible with the weak field view (because there are then in
fact no vacuum regions, such as represented in the Lindquist-Wheeler
type models). This may or may not be the case.

In \cite{Bucetal09} it is shown that the second derivatives can be of order one in the situation given by the other numbers for metrical perturbations. Curvature is thus important and not a perturbation of a flat model; but it is the curvature that drives the backreaction effect. The degree to which a suitable linearised approach is acceptable as a model of genuinely inhomogenous regions thus remains open to debate, particularly in the case of a linearised treatment on a flat background, where the curvature remains small. 

\subsubsection{The gauge issue}
Finally underlying this all is the gauge issue: to what degree are the results dependent on the choice of how the background metric is mapped into the more realistic model?  One can after all always find a gauge where the density perturbation $\delta\rho$ is zero \cite{EllBru89}. The key is to find a gauge invariant formalism to tackle the problem  - if that is possible \cite{EllMat95}. The major attempt to tackle this so far is by Gasperini, Marozzi, and Venziano \cite{GasMarVen09,GasMarVen10}. This has not yet however led to specific conclusions about cosmological acceleration. This issue is related to the complexities of appopriately defining the background spacetime \cite{EllSto87,KolMatRio10}.\\ 

The overall conclusion is that while it may at first seem rather unlikely that dynamical backreaction is of significance in the late universe, there are some unresolved questions, so that one should keep an open mind. The issue is debated in some of the following papers in this special issue. Furthermore it may be important in the early universe: for example Mukhanov {\it et al} have shown that the back reaction of cosmological perturbations on the background can become important already at energies below the self-reproduction scale in inflationary universe scenarios \cite{MukAbrBra97}. However I will not discuss that context here.


\section{Optical effects of local inhomogeneity}

Small scale inhomogeneity can have significant effects on the
propagation of photons in a lumpy universe, with potentially
important effects on observations. There are three issues here.

\subsection{Redshift effects} Firstly, inhomogeneities can affect redshifts, as for example in the
Rees-Sciama effect \cite{ReeSci67} where CMB photons falling into a
time-dependent gravitational potential well experience an overall
change in redshift because they climb out of a different shaped well
than when they fell in. Also if light is emitted from a source
within a potential well, it will be redshifted as it climbs out;
this effect lies behind the `timescape cosmology' proposal of
Wiltshire \cite{Wil08}, who points out that the associated time
dilation 
effect is cumulative over the history of the source.

\subsection{Area distance effects} Secondly inhomogeneities can affect area distances, which underlie
the apparent angular diameter, and hence apparent luminosity, of
images \cite{Ell71}. The key point is the difference between Ricci
focusing and Weyl focusing, as emphasized by Bertotti \cite{Ber66}.
The focussing of an irrotational bundle of null geodesics with
tangent vector $K^a$ is given by
\begin{equation}\label{null1}
d\hat{\theta}/dv = -R_{ab}K^aK^b - 2\hat{\sigma}^2 - \hat{\theta}^2
\end{equation}
\begin{equation}\label{null2}
d \hat{\sigma}_{mn}/dv  = - E_{mn}
\end{equation}
where $\hat{\theta}$ is the expansion and $\hat{\sigma}$ the shear of the
null rays, $R_{ab}$ is the Ricci tensor, determined pointwise by the
matter distribution, and $E_{ab}$ the electric part of the Weyl
tensor, determined non-locally by matter elsewhere.

In the case of Robertson-Walker observations, there is zero Weyl
tensor and a non-zero Ricci tensor, so (\ref{null1},\ref{null2})
 become
\begin{equation}\label{null3}
d\hat{\theta}/dv = -R_{ab}K^aK^b  - \hat{\theta}^2
\end{equation}
\begin{equation}\label{null4}
d\hat{\sigma}_{mn}/dv  = 0
\end{equation}
which are the standard equations underlying observations in a FLRW
model. Actual observations however are the opposite: photons travel
through empty space (on small scales), described by zero Ricci
tensor and non-zero Weyl tensor: so (\ref{null1},\ref{null2}) become
\begin{equation}\label{null5}
d\hat{\theta}/dv =  - 2\hat{\sigma}^2 - \hat{\theta}^2
\end{equation}
\begin{equation}\label{null6}
d \hat{\sigma}_{mn}/dv  = - E_{mn}
\end{equation}
This averages out to FLRW equations when averaged over whole sky,
which is not obvious! This does not follow from energy conservation
\emph{per se}, but rather  depends on how area distances average out
over the sky. But supernova observations are preferentially made in
directions where there is no matter in between to interfere with the
observations; hence area distances, and so cosmological
observations, will be different in this case.

The usual way of handling this is to use the Dyer-Roeder (DR)
equation \cite{DyeRoe73,DyeRoe74,EhlSchFal86}, that takes matter
into account but not shear, because the shear enters the focusing
equation quadratically, and so is neglible if shear is small.  Thus
the DR equation takes in to account only the Ricci focusing due to a
specified fraction $f(v)$ of the uniform density of matter in the
universe :
\begin{equation}\label{null7}
d\hat{\theta}/dv = - f(v) R_{ab}K^aK^b  - \hat{\theta}^2.
\end{equation}
When $f=1$ one has the FLRW result; when $f = 0$ one has photons
travelling through vacuum regions in the clumpy universe.

How this works out depends on how dark matter is clustered, which
differs on different scales. The Dyer-Roeder approximation is good
if the Weyl focusing term (causing gravitational lensing) can always
be neglected in this way; this needs investigation in the light of
the expected clustering pattern; many examples are given by Mortsell  \cite{Mor02}, 
showing the effect is potentially significant, and analytic forms by Kantowski \cite{Kan03}. When this approximation is valid, the outcome
depends crucially on what fraction of the overall cosmic density
(baryonic and non-baryonic) occurs in a smooth form along the line
of sight on different scales. Note that on some angular scales the
clumping experienced along the line of sight will be partially
compensated, in that each void (a low density region on the line of
sight) will be matched by a wall (a high density region) so that the
overall density is the same as the background density. However they
will not exactly compensate because 3-dimensionally compensated
voids do not reduce to a 1-dimensionally compensated distribution of
matter along the line of sight \cite{Bol10}. It will also have some impact
on the CMB observations \cite{Bol11}.

One can investigate these effects in non-linear models. How it works
out in Swiss cheese models is investigated {\it inter alia} in
\cite{Kan95,Kan98}, confirming there can indeed be a significant
effect. The case of observations in a Wheeler-Lindquist type model
is investigated in \cite{CliFer10,Cli10}.

The key issue is how empty the voids really are, from supergalactic
scales down to the "vacuum" regions in the solar system. There are
some galaxies in the large scale voids, but are they imbedded in an
intergalactic gas of baryons and CDM? If so what what fraction is
its density of the global average density of the model (when
smoothed on the largest scales)? The answer does not seem to be
known: but the outcome depends crucially on these figures. On the
small scales relevant to the supernova observations, one may expect
mostly empty space, except perhaps for CDM left over from structure
formation; but it is unclear what the relevant fractional density is
on these scales

\subsection{Afine parameter effects}
Finally, there are effects that arise through altering the relation $z(v)$ between the affine parameter $z$ 
and the redshift $z$. These effects have been little studied. However it is noteworthy that it is only through this relation
that the cosmolgical constant can affect observation relatons such as the area-distance redshift relation ($\Lambda$ does not
explicitly enter the null Raychaudhuri relation (\ref{null1})). Thus this may well be interesting to investigate.\\

Overall these effects are indeed likely to be significant: that is, they may be significant enough to appreciably 
affect the parameter values of the concordance model of cosmology \cite{Jaretal10,Peretal10}. How this works
out is crucially dependent on how matter is distributed on small
scales, and how empty the voids really are. This is an important
area for investigation, and is discussed by Mattsson.

\section{Spatial homogeneity?}

So far I have considered the effect of local inhomogeneities on
global dynamics and observations; where "local" means sufficiently
small that we can claim that overall the Copernican principle - the
claim that the universe is the same everywhere - still holds when we
coarse-grain on large enough scales. The further issue of interest
is whether this is in fact the case: might it be that the Copernican
principle does not hold, so the FLRW models are in fact misleading
models of the large scale geometry of the visible region of the
universe?

The Cosmological Principle was introduced by Milne 1930's, and then
formalized in a technical sense by Robertson and Walker. It was the
foundation of cosmology in the 1960s to 1980s, see Bondi
\cite{Bon60} and Weinberg \cite{Wei72}. But it is an \emph{a priori}
philosophical principle. It produces world models that work - namely
the standard models of cosmology. But is it true? Can it be tested?
Maybe there are inhomogeneous models that would fit the observations
as well - or even better.

\subsection{The argument for homogeneity}
It is not obvious the universe is spatially homogeneous
\cite{Ell75,Ell79}. We can directly observe isotropy, but not
homogeneity, firstly because we effectively observe the universe
from one space-time point, and secondly because when undertaking
astronomical observations, the finite speed of light inextricably
mixes spatial distance with time.

Arguments for homogeneity are discussed in \cite{Ell06}. Direct
determination of homogeneity from number counts is in principle
possible, but fails in practice because of the look-back time
necessarily associated with all cosmological observations: we cannot
uniquely separate out spatial inhomogeneity from a time evolution of
sources \cite{Ell75,MusHelEll99}. Similarly in principle an
observational verification of the Mattig magnitude-redshift relation
for galaxies in FLRW models \cite{San61,Ell71} (or its
generalization to non-zero $\Lambda$) would suffice \cite{Elletal
85}. This in-principle direct determinations of
homogeneity depends on being precisely fit
by FLRW data functions, and does not depend on
observations by other fundamental observers. But again this is not practicable. So how can one proceed?

The high degree of isotropy of astronomical observations (averaged
on a large enough scale) suggests an observational basis for the
assumption of spatial homogeneity. Indeed a universe which is
isotropic everywhere is necessarily a FLRW model (Walker
\cite{Wal35}, Ehlers \cite{Ehl60}). But we can't check if this is
true or not: it's an assumption, because we can only test isotropy
where we are. However we can attain a weaker version of the Walker
result: Ehlers, Geren and Sachs \cite{EhlGerSac86} proved the EGS
theorem, that isotropy everywhere of the CBR only is sufficient to
prove a FLRW geometry, if the universe is expanding. This result has been strengthened 
even further through generalisations of the EGS theorem 
to almost isotropy and to  models with matter and dark energy \cite{StoMaaEll95,ClaMaa10}. 
This provides a stronger motivation for spatial homogeneity, but until recently still relied on an
untested philosophical assumption: addition of a Copernican
Principle, assuming that we are not in a special position in the
universe, so everyone else will also see isotropic background
radiation. The result then follows. However it is now known that this assumption is indeed at least partly testable
via measurements of CMB spectrum distortions, as will be discussed below.

There are a number of other observational tests of the
Copernican principle that are now possible, because of observational
improvements in the past decade. Before coming to them, I will first
discuss the inhomogeneous models that make this an interesting
possibility.

\subsection{Large scale inhomogeneity?}
The proposal that inhomogeneous models can explain the supernova
observations without any dark energy is discussed by
C\'{e}l\'{e}rier  \cite{Cel07} and Tanimoto \& T. Nambu
\cite{TanNam07}. The idea is that there is a large scale
inhomogeneity of the observable universe such as that described by
the Lema\^{\i}tre-Tolman-Bondi (LTB) pressure-free spherically
symmetric models (\cite{Bon47}, see also see \cite{Kra97,BKHC2009,Hel09}), and we are near the centre of a void. The LTB
models have comoving coordinates

$$ds^2 = - dt^2 + B^2(r,t) + A^2(r,t)(d\theta^2+\sin^2 \theta
d\phi^2)$$ where

$$B^2(r,t) = A'(r,t)^2 (1-k(r))^{-1}$$
and the evolution equation is

$$(\dot{A}/A)^2 = F(r)/A^3 + 8pG\rho_\Lambda/3 - k(r)/A^2$$
with the energy density given by $F' (A'A^2)^{-1} = 8pG\rho_M$.
There are two arbitrary functions of the spatial coordinate $r$:
namely $k(r)$ (curvature) and $F(r)$ (matter).
That this freedom enables us to fit the supernova observations with
no dark energy or other exotic physics is a consequence of a theorem
proved by Mustapha \emph{et al} \cite{MusHelEll99}, updated in
\cite{LuHel07,McCHel08}. One can also fit the basic nucleosynthesis
data and CBR observations because they refer to much larger values
of $r$ see e.g. Alexander \emph{et al} \cite{aletal07}. The key
point is that different scales are probed by different astronomical
observations and can in principle all be fitted by adjusting the
free spatial functions at different distances. One can  also use Baryon Acoustic
Oscillation (BAO) measurements to estimate distances \cite{ClaReg10}; but note that to calculate the CBR and BAO results 
with precision,  one must use the LTB perturbation theory \cite{ClaCliFeb09}, not the theory of FLRW perturbations.

A typical observationally viable model is one in which we live
roughly centrally (within 10\% of the central position) in a large
void with a compensated underdense region stretching to $z \simeq
0.08$ with $\delta \rho/\rho \simeq -0.4$ and size $160/h$ Mpc to
$250/h$ Mpc, a jump in the Hubble constant of about $1.20$ at that
distance, and no dark energy or quintessence field
(\cite{bisetal06,AlAmGr06,yooetal08}). Actually you don't need a
void to explain the observations; more general models can do the job
\cite{Hel09,CeBoKr10}. One can also use the more complex Szekeres
universes to obtain observationally viable models \cite{ishetal07}.

One ends up with a degeneracy: both FLRW and LTB models can explain
the basic cosmological observations, as was confirmed for example by the SDSS team \cite{Soletal09}. 
One needs more detailed
modeling to distinguish which is the better model when precision
cosmological observations are taken into account. Before I turn to
these tests, some theoretical objections must be faced.

\subsection{Dynamical origins and Probability}
Given that we can fit the observations by such a model, is there a
plausible dynamic scenario for them? Because evolution along
individual world lines in such models is governed by the Friedmann
equation, inflation followed by a Hot Big Bang era can have the same
basic dynamics as in the standard model, but with position dependent
parameters. One argument for homogeneity is that inflation creates a
high degree of uniformity, and in the subsequent cosmic evolution,
perturbations can only grow to a certain size. Above that scale, we
should have the inflation-created uniformity. But that depends on
the details supposed for the inflationary epoch. If there are
multiple inflaton fields and appropriate inflationary potential and
initial conditions, then it should certainly be possible to arrive
at an inhomogeneous a situation, for example multi-stream inflation
\cite{AfsSloYan10} gives such a mechanism. This involves two
inflaton fields, a hill in the potential, and tunneling between
different paths from initial to final states, resulting in different
numbers of e-foldings in different places. This mechanism can create
large over or under densities of the kind envisaged here.

Many dismiss these models on probability grounds: "t is improbable
we are near the centre of such a model." But there is always
improbability in cosmology: we can shift it around, but it is always
there. Three comments are in order. First, there simply is no proof
the universe is probable; that is a philosophical assumption, which
may not be true. Secondly, a study by Linde \emph{et al}
\cite{linetal95} shows that (given a particular choice of measure)
this kind of inhomogeneity actually is a probable outcome of
inflationary theory, with ourselves being located near the centre.
 And thirdly, Boljeko and Sussman argue \cite{BolSus10} that the
 problem of improbability is ameliorated if
one has for example a Szekeres rather than LTB solution.

Overall, one cannot simply dismiss such models out of hand.
Philosophical opinions and probability arguments will have to give
way to the results of observational testing of these models.

\subsection{Observational tests of spatial homogeneity} Given that
we can both find inhomogeneous models to reproduce the observations
without any exotic energy, as well as homogeneous models with some
form of dark energy that explain the same observations, can we
distinguish between the two? Ideally we need a model-independent
test: is a RW geometry the correct metric for the observed universe
region, irrespective of assumptions about the dynamics and matter
content? Four kinds of tests are possible.

\subsubsection{Behaviour near origin} The
universe must not have a geometric cusp at the origin, as this
implies a singularity there. Thus it has been claimed there are
centrality conditions that must be fulfilled in the inhomogeneous
models (Vanderveld et al \cite{vanetal06}). The distance modulus
behaves as $\Delta dm(z) = - (5/2)q_0z$ in standard $\Lambda$CDM
models, but if this were true in a LTB void model without
$\Lambda$ this has been said to imply a singularity (Clifton
\emph{et al} \cite{clietal08}); observational tests of this
requirement will be available from intermediate redshift
supernovae in the future. However \cite{Febetal10} and
\cite{Kraetal10} show this is not a real issue.

\subsubsection{Area distance versus Hubble parameter}
Measures of the area distance and Hubble parameter as a function of
redshift can give a direct test of spatial homogeneity. There are
two geometric effects on distance measurements: curvature $\Omega_k$
bends null geodesics, expansion $H(z)$ changes radial distances.
In RW geometries, we can combine the Hubble rate and distance data
to find the curvature today:
$$\Omega_k = \frac{[H(z)D'(z)]^2-1}{[H_0D(z)]^2} $$
This relation is independent of all other cosmological parameters,
including dark energy model and theory of gravity. It can be used at
single redshift to determine $\Omega_k$, but must give the same
result for all redshifts. The important result of Clarkson \emph{et
al} (\cite{claetal08}) is that since $\Omega_k$ is independent of
$z$, we can differentiate to get the consistency relation
$$ C(z) := 1 + H^2(DD''-D'^2) + HH'DD' =0, $$
which depends only on a RW geometry: it is independent of curvature,
dark energy, nature of matter, and theory of gravity. Thus it gives
the desired consistency test for spatial homogeneity. In realistic
models we should expect $C(z) \simeq 10^{-5}$, reflecting
perturbations about the RW model related to structure formation.
Errors may be estimated from a series expansion

$$C(z) = \left[q_0^{(D)}-q_0^{(H)}\right]z + O(z^2)$$
where $q_0^{(D)}$ is measured from distance data and $q_0^{(H)}$
from the Hubble parameter. It is simplest to measure $H(z)$ from BAO
data. It is only as difficult carrying out this test as carrying out
dark energy measurements of $w(z)$ from Hubble data, which requires
$H'(z)$ from distance measurements or the second derivative
$D''(z)$. 
Another promising approach is to use the time
drift of cosmological redshifts as a way of determining these
functions \cite{UzaClaEll08}. An analysis of how well the time drift
of redshift $\dot {z}$ can distinguish an LTB model from a FLRW
model is given in \cite{Dunetal10}.

This is the simplest direct test of spatial homogeneity, and its
implementation should be regarded as a high priority: for if it
confirms spatial homogeneity, that reinforces the evidence for the
standard view in a satisfying way; but if it does not, it has the
possibility of undermining the entire project of searching for a
physical form of dark energy.

In the future, the same measurements can potentially be carried
out by gravitational wave observations of black hole binary
mergers \cite{Hugetal01,JonGooMor07,PetBabSes11}.

\subsubsection{The CMB Spectrum: Verifying the EGS Theorem conditions}
The peaks in the CMB anisotropy power spectrum can be adequately
accommodated in the LTB family of models \cite{ClaReg10}. 
The key further point is that one can use scattered CMB photons to check CMB isotropy at points
away from the origin (Goodman \cite{goo95}; Caldwell and Stebbins
\cite{calste07}), thus checking some of the conditions required by the EGS
theorem. 

If the CMB radiation is anisotropic around distant observers (as will be true in inhomogeneous models), then 
the Sunyaev-Zeldovich scattered photons will cause a distorted spectrum CMB spectrum, as anisotropy of the CMB out
there will cause a mixing of temperatures in the scattered photons. Such anisotropy can arise in two ways  \cite{calste07}. Firstly, the 
Kinematic SZ effect occurs due to via relative motion between matter and the CMB at distant points.  Gradients in the void gravitational potential causes gas to move relative to CMB frame, hence there will be a CMB dipole out there. This violates the EGS conditions, and scattering mixes these temperatures, causing a spectral distortion. Secondly, potential wells cause anisotropy due to gravitational redshift effects. If some photons originate inside the void and others outside, this again causes a locally anisotropic CMB out there, and SZ scattering compares potentials at the two locations.

It has recently been claimed by two groups that such CMB observations
disprove inhomogeneity \cite{MosZibSco10,ZhaSte10}, but counter
claims \cite{ClaReg10} give specific models where the CMB
observations are acceptably accounted for.

The problem seems to be firstly that the papers
\cite{MosZibSco10,ZhaSte10} refer to restricted families of LTB
Models, which have to be generalized to include radiation effects
in order to handle the CMB observations; the radiation and the
matter may not be comoving \cite{ClaReg10}. Also if one only
considers LTB models with fixed bang-time, one has removed half
the freedom of the LTB models; it is then hardly surprising if
fitting the observations is difficult. Generic analysis should
allow varying bang time. Secondly, these are not self consistent
studies, as they use FLRW perturbation theory to study structure
formation in LTB models. One needs to use LTB perturbation theory
\cite{ClaCliFeb09} to get consistent results.

Future work of interest here will be to check to what degree such tests can verify the full requirements 
of the extended versions of the EGS theorem discussed by Clarkson and Maartens \cite{ClaMaa10}. Can they fully test 
the needed anisotropy requirements for one of the extended versions of the EGS results, or do they only serve as partial checks 
of the needed conditions, because they only check mixing of lower order CMB multipoles?

\subsubsection{Thermal History based tests}
 If the kinds of structures that occur in distant regions are similar
 to those nearby, that indicates that the thermal histories leading to
 the existence of those structures must have been the same; and this
 suggests that the universe must have been spatially homogeneous at the
 relevant early times - which will imply it is homogeneous today. This is
 the \emph{Postulate of Uniform Thermal Histories} (`PUTH')
 \cite{BonEll86}. Conversely, if the kinds of objects that have come into
 being far away look different from those nearby, this indicates
 spatial inhomogeneity.

In principle this can be applied for example to studies of galaxies
and and Large Scale Structure; this has not yet been formally done.
However a present application is to element abundances. There are
now claims of some anomalies in the abundance of lithium with
distance (see e.g. \cite{Koretal06}). Regis and Clarkson
\cite{RegCla10} show that this can be taken as indirect evidence for
spatial inhomogeneity.\\

Observations in inhomogeneous models are discussed in the papers by
Zibin by V. Marra and A. Notari.


\section{Conclusion}

An implicit averaging is effectively at the foundation of how the
standard model deals with matter and structure formation, while
being uniform on large scales. The problem of averaging is far from
solved - but it is a problem that will not go away. Smaller scale
inhomogeneities may possibly cause observable effects through
dynamical backreaction, but this is a controversial suggestion.
Ultimately, we probably need a general relativistic simulation of
structure formation to resolve the issue of averaging. However, such
inhomogeneities certainly can significantly affect the observational
determination of the parameters of the concordance cosmological
model. Whether this is the case or not depends on the detailed
nature of clustering of dark matter, on small scales, in the
universe, which is not known at present.

Additionally, we must take seriously the idea that the
acceleration apparently indicated by supernova data could be due to
large scale inhomogeneity with no dark energy. Observational tests
of the latter possibility are as important as pursuing the dark
energy (exotic physics) option in a homogeneous universe.
Theoretical prejudices as to the universe's geometry, and our
place in it, must bow to such observational tests. Precisely
because of the foundational nature of the Copernican Principle for
standard cosmology, we need to fully check this foundation. And
one must emphasize here that standard CMB anisotropy studies do
not \emph{prove} the Copernican principle: they \emph{assume} it
at the start.

Whatever the outcome of these studies, the point remains that
inhomogeneity is a critical topic in cosmology. Simplified models of
inhomogeneity such as LTB models, where we can actually calculate
dynamics and predict observational relations, are an important part
of the necessity to probe every aspect of the standard model, as are
studies of the nature of the backreaction effect and the effects of
inhomogeneities on observations.

\subsection{To be done}

To complete our understanding of these issue, inter alia we need to
\begin{itemize}
  \item develop a general
relativistic simulation of structure formation
  \item  develop perturbation
studies of the LTB models, and hence CMB anisotropies and LSS
observations, in a self-consistent way
  \item develop the PUTH approach \cite{BonEll86} for galaxies and LSS
\item Use observations and simulations to characterize in detail the
DM inhomogeneity on small scales, and find out to what degree  it
clusters with baryons on these scales
\item hence to characterise in detail the DM and baryonic IGM (Inter-Galactic Medium) that may
permeate the `voids' in the cosmic web, at different scales
\item along with determining homogeneity,  we really need to
determine the smallest length scale on which the universe is almost-FLRW, if
indeed it is almost-FLRW on large scales. This is related to the possibility
that some of the data we use for determining the cosmology may not be probing
almost-FLRW scales of the universe -- e. g. may not be probing the Hubble flow.
\end{itemize}
Finally on should remember that the issues mentioned here are not mutually exclusive.  
If we do live in a Hubble scale inhomogeneity, the universe is additionally inhmogeneous on smaller scales. Hence 
the eventual aim must be to investigate the combination of all these effects.


\section*{Acknowledgements}

I thank Bill Stoeger, Thomas Buchert, Chris Clarkson, Jean-Philippe
Uzan, Roy Maartens, Julien Larena, and Obinna Umeh for helpful
discussions on these topics, and Charles Hellaby, Thomas Buchert, Bill Stoeger,
and Alan Coley for useful comments on previous versions of this text. This work has been supported by
the South African National Research Foundation (NRF) and University
of Cape Town (UCT).



\begin{thebibliography}{99}

\bibitem{AfsSloYan10} N Afshordi, A Slosar, Y Wang (2010) ``A Theory of a
Spot '' (arXiv:1006.5021)

\bibitem{aletal07} S Alexander, T Biswas, A Notari, A., and D Vaid
(2009) "Local void vs dark energy: confrontation with WMAP and Type
IA supernovae"  JCAP 0909:025 [arXiv:0712.0370].

 \bibitem{AlAmGr06}
      H. Alnes, M. Amarzguioui \& O. Gr\o n,
      ``Inhomogeneous Alternative to Dark Energy?''      {\it Phys. Rev. D} {\bf 78}, 083519 (2006),
 [arXiv:astro-ph/0512006]

\bibitem{triangle}
Neta A. Bahcall, Jeremiah P. Ostriker, Saul Perlmutter and Paul J.
Steinhardt (1999) ``The Cosmic Triangle: Revealing the State of the
Universe'' \emph{Science} \textbf{284}: 1481-1488 [	arXiv:astro-ph/9906463v4]

\bibitem{Bauetal10} Daniel Baumann, Alberto Nicolis, Leonardo Senatore,
and Matias Zaldarriaga (2010) ''Cosmological Non-Linearities as an
Effective Fluid''. arXiv:1004.2488.

\bibitem{GA3}
Bennett \emph{et al}. (2003) ''First Year Wilkinson Microwave
Anisotropy Probe (WMAP) Observations: Preliminary Maps and Basic
Results'' Astrophys.J.Suppl.148:1,2003 [astro-ph/0302207].

\bibitem{Ber66} B Bertotti (1966) ``The luminosity of distant galaxies"
{\it Proceedings of the Royal Society of London}. Series A, Mathematical
and Physical Sciences {\bf 294}: 195-207

\bibitem{BilFut91} S. Bildhauer and T. Futamase, (1991) ``The age problem in inhomogeneous universes'' {\it Gen. Rel. Grav}. {\bf 23}, 1251.

\bibitem{bisetal06}
 T Biswas, R Monsouri, R., and A Notari (2007)
 "Nonlinear Structure Formation and `Apparent' Acceleration: an Investigation"
  JCAP 12, 017 [astro-ph/0606703].

\bibitem{Bol10} Krzysztof Bolejko (2010) ``Weak lensing and the Dyer-Roeder
approximation'': arXiv:1011.3876.

\bibitem{Bol11}
Krzysztof Bolejko (2011)
``The effect of inhomogeneities on the distance to the last scattering surface and the accuracy of the CMB analysis'':
arXiv:1101.3338. 


      \bibitem{BKHC2009}
      K. Bolejko, A. Krasi\'nski, C. Hellaby \& M-N. C\'el\'erier
      (2009)
      {\it Structures in the Universe by Exact Methods --- Formation,
Evolution, Interactions},
      (Cambridge: Cambridge University Press).

\bibitem{BolSus10} Krzysztof Bolejko, Roberto A. Sussman (2011) ``Cosmic
spherical void via coarse-graining and averaging non-spherical
structures'' {\it Physics Letters} {\bf  B 697}: 265-270 [arXiv:1008.3420].

\bibitem{Bon47}
H Bondi (1947) ``Spherically symmetric models in General Relativity'' {\it Mon Not Roy Ast Soc} {\bf 107}: 410. Reprinted in {\it Gen Rel Grav}
{\bf 31}, 1783-1805 (1999(,

\bibitem{Bon60} H Bondi (1960), \emph{Cosmology}. (Cambridge University Press,
Cambridge.

\bibitem{BonEll86} W B Bonnor and G F R Ellis (1986), ``Observational
homogeneity of the universe". \emph{Mon Not Roy Ast Soc}
\textbf{218}, 605-614.

\bibitem{BriHar64} Dieter R. Brill and James B. Hartle (1964). ``Method of the
Self-Consistent Field in General Relativity and its Application to
the Gravitational Geon'' \emph{Phys. Rev}. \textbf{135}, B271-- B278.

\bibitem{BucEhl97} T Buchert, and J Ehlers (1997) ``Averaging inhomogeneous Newtonian
cosmologies.'' {\it Astronomy and Astrophysics} {\bf 320}: 1-7.

\bibitem{Buc99} T Buchert (2000) ``On Average Properties of Inhomogeneous Fluids
in General Relativity: Dust Cosmologies''
  GRG 32: 105-126 [arXiv:gr-qc/9906015]

\bibitem{Buc08} T Buchert (2008) ``Dark Energy
from structure: a status report'' {\it Gen. Rel. Grav}. {\bf 40}, 467-527  [arXiv:0707.2153].

\bibitem{Bucetal00} T Buchert, M Kerscher, and C Sicka
(2000)  ``Back reaction of inhomogeneities on the expansion: The
evolution of cosmological parameters'' \emph{Physical Review}
\textbf{D: 62}: 043525 (arXiv:astro-ph/9912347)

\bibitem{BucCar01}   T. Buchert and M. Carfora (2002), ``Matter seen at many scales:
the geometry of averaging in relativistic cosmology''. In
{\it General Relativity, Cosmology, and Gravitational Lensing}, G. Marmo, C. Rubano and P. Scudellaro (eds.), 
Napoli Series on Physics and Astrophysics, Bibliopolis, Naples, 29-44 [arXiv:gr-qc/0101070].

\bibitem{BucCar02}  T. Buchert and M. Carfora (2002), `Regional averaging and scaling in relativistic
cosmology', {\it Class. Quant. Grav}. {\bf 19}, 6109-6145 [arXiv:gr-qc/0210037].

\bibitem{BucCar03}   T. Buchert and M. Carfora (2003), ``The Cosmic Quartet - Cosmological Parameters of a
Smoothed Inhomogeneous Spacetime''. Proc. 12th JGRG conference,
Tokyo 2002, Shibata M. et al. (eds.), 157-161
[arXiv:astro-ph/0312621]

\bibitem{Bucetal09} T. Buchert, G.F.R.
Ellis and H. v. Elst, (2009) ``Geometrical order-of-magnitude
estimates for spatial curvature in realistic models of the
Universe'' \emph{Gen. Rel. Grav}. \textbf{41}, 2017-2030 [arXiv:0906.0134].

\bibitem{BucCar08} Thomas Buchert and Mauro Carfora (2008) ``On the curvature of
the present-day Universe'' \emph{Class.Quant.Grav}. \textbf{25}:
195001 [arXiv:0803.1401v2]

\bibitem{calste07} R Caldwell and  A Stebbins, A. (2008) "A Test of the Copernican Principle" \emph{Phys Rev Lett}. \textbf{100},
191302  [arXiv:0711.3459].

\bibitem{Car10} Mauro Carfora (2010)
``Ricci Flow Conjugated Initial Data Sets for Einstein Equations'':
 arXiv:1006.1500. 

\bibitem{CarPio95} M Carfora and K Piotrkowska (1995) `The
renormalistion group approach to relativistic cosmology'' \emph{Phys
Rev} \textbf{D52}:4393 [ arXiv:gr-qc/9502021].

\bibitem{Cel07} M-N. C\'{e}l\'{e}rier (2007) ``The Accelerated Expansion of the Universe Challenged 
by an Effect of the Inhomogeneities. A Review'' \emph{New Advances in Physics} \textbf{1}, 29
[astro-ph/0702416].

 \bibitem{CeBoKr10}
      M-N. C\'{e}l\'{e}rier, K. Bolejko \& A. Krasinski  (2010)
      ``A (giant) void is not mandatory to explain away dark energy with
a Lema\^{\i}tre - Tolman model''.{\it Astron. Astrophys.\/} {\bf
518}, A21[ arXiv:0906.0905].

\bibitem{ClaAnaLar09} Chris Clarkson, Kishore Ananda, Julien Larena (2009)
``The influence of structure formation on the cosmic expansion''
{\it Phys.Rev}. {\bf D80}: 083525 [arXiv:0907.3377]

\bibitem{claetal08} C Clarkson, B Bassett, B., and T. H-C. Lu,  (2008). "A
general test of the Copernican Principle". {\it Phys Rev Lett} {\bf 101},
011301 [arXiv:0712.3457].

\bibitem{ClaCliFeb09} Chris Clarkson, Timothy Clifton and  Sean February (2009) ``Perturbation
Theory in Lemaitre-Tolman-Bondi Cosmology'' \emph{JCAP} \textbf{06}:
025 [arXiv:0903.5040]


\bibitem{Claetal11}
Chris Clarkson, G.F.R Ellis, Julien Larena and Obinna Umeh (2011)
``Averaging and backreaction in cosmology: Is it important?'' To appear, {\it Rep Prog Phys}.

\bibitem{ClaMaa10} C Clarkson and R Maartens (2010) ``Inhomogeneity and the
foundations of concordance cosmology''\emph{ Class. Quantum Grav}.
\textbf{27} 124008 [arXiv:1005.2165]

\bibitem{ClaReg10} Chris Clarkson, Marco Regis  (2011) ``The Cosmic
Microwave Background in an Inhomogeneous Universe - why void models
of dark energy are only weakly constrained by the CMB'' 	JCAP02: 013
[arXiv:1007.3443].

\bibitem{Cli10} Timothy Clifton (2010) ``Cosmology Without
Averaging'' [arXiv:1005.0788]

\bibitem{clietal08} T Clifton, P G Ferreira, and K Land (2008). "Living
in a Void: Testing the Copernican Principle with Distant
Supernovae", \emph{Phys Rev Lett}, \textbf{101}, 131302
[arXiv:0807.1443].

\bibitem{CliFer10} Timothy Clifton and Pedro G.
Ferreira (2009) ``Archipelagian Cosmology: Dynamics and Observables
in a Universe with Discretized Matter''
\emph{Phys.Rev}.\textbf{D80}: 103503 [arXiv:0907.4109]

\bibitem{Col10}
A.A. Coley (2010)
``Averaging in cosmological models using scalars'' {\it Class.Quant.Grav.} {\it 27}: 245017  [arXiv:0908.4281].

\bibitem{ColPelZal05} A. A. Coley, N. Pelavas, and R. M. Zalaletdinov
(2005) ``Cosmological Solutions in Macroscopic Gravity'' \emph{Phys.
Rev. Lett}. \textbf{95}, 151102 [ arXiv:gr-qc/0504115 ].

\bibitem{Dod03}
S Dodelson (2003) \emph{Modern Cosmology} (Elsevier).

\bibitem{Dunetal10}
P. Dunsby, N. Goheer, B. Osano, and J.-P. Uzan (2010) ``How close
can an inhomogeneous universe mimic the concordance model?'' JCAP
1006 (2010) 017 [arXiv:1002.2397].

\bibitem{DyeRoe73} C C Dyer and R C Roeder (1973) ``Distance-Redshift
Relations for Universes with Some Intergalactic Medium''
\emph{Astrophysical Journal} \textbf{180}, L31.

\bibitem{DyeRoe74} C C Dyer and R C Roeder  (1974) ``Observations in
Locally Inhomogeneous Cosmological Models'' \emph{Astrophysical
Journal,} \textbf{189}, 167-176.

\bibitem{Ehl60} J Ehlers (1960), ``Contributions to the relativistic mechanics of
continuous media". \emph{Abh Mainz Akad Wiss u Lit (Math Nat Kl)}
(1960); reprinted in English, \emph{Gen Rel Grav} \textbf{25},
1225-1266 (1993).

\bibitem{EhlGerSac86} J Ehlers, P Geren and R K Sachs  (1968), ``Isotropic solutions
of the Einstein-Liouville equations". \emph{J Math Phys} \textbf{9},
1344-1349.

\bibitem{EhlSchFal86} J Ehlers, P Schneider and E E Falco (1992)
\emph{Gravitational Lenses}, (New York: Springer, 1992).

\bibitem{EinStr45}
A Einstein, A. and E G Straus  (1945), Rev. Mod. Phys. {\bf17}, 120.

\bibitem{EinStr45a}
A Einstein and E G Straus  (1945), Rev. Mod. Phys. {\bf18}, 148.

\bibitem{Ell71} G F R Ellis (1971): ``Relativistic Cosmology'', in \emph{General
Relativity and Cosmology}, Proceedings of the XLkII Enrico Fermi
Summer School, Ed. R K Sachs, (New York: Academic Press).

\bibitem{Ell75} G F R Ellis   (1975), ``Cosmology and Verifiability". \emph{Qu Journ
Roy Ast Soc} \textbf{16}, 245-264.

\bibitem{Ell79} G F R Ellis   (1979) ``The homogeneity of the universe". \emph{Gen
Rel Grav} \textbf{11}, 281-289.

\bibitem{Ell84} G F R Ellis   (1984): ``Relativistic cosmology: its nature,
aims and problems". In \emph{General Relativity and Gravitation}, Ed
B Bertotti et al (Reidel), 215.

\bibitem{Ell06}
G F R Ellis   (2006). "Issue in the Philosophy of cosmology". In
\emph{Handbook in Philosophy of Physics}, Ed J Butterfield and J
Earman. Dordrecht: Elsevier, 1183 [astro-ph/0602280].

\bibitem{EllBuc05} G F R Ellis and T R Buchert (2005): "The universe seen at different
scales". Physics Letter A 347: 38-46

\bibitem{EllBru89} G. F. R. Ellis and M Bruni (1989) ``Covariant and gauge-invariant approach to cosmological
density fluctuations'' \emph{Phys. Rev. D} \textbf{40}, 1804–1818

\bibitem{EllMaaMac11} G F R Ellis, R Maartens
and M A H MacCallum (2011) \emph{Relativistic Cosmology} (Cambridge:
Cambridge University Press)

\bibitem{EllMat95}
G F R Ellis and D R Matravers (1995): ``General Covariance in General Relativity''.   {\it  Gen Rel Grav} {\bf  27}, 777.   

\bibitem{Elletal 85}
G F R Ellis, S D Nel, W Stoeger, R Maartens, and A P Whitman: `Ideal
Observational Cosmology". Phys Reports 124, Nos 5 and 6, 315-417
(1985).

\bibitem{EllSto87} G F R Ellis and W R Stoeger (1987): ``The
Fitting Problem in Cosmology". Class Qu Grav 4, 1679-1690 .

\bibitem{Febetal10}
Sean February, Julien Larena, Mathew Smith, Chris Clarkson
``Rendering Dark Energy Void" (2010) \emph{Mon. Not. Royal Astr.
Soc}. \textbf{405}, 2231 [arXiv:0909.1479v2]

\bibitem{web}
J. E. Forero-Romero, Y. Hoffman, S. Gottl\"{o}ber, A. Klypin, G.
 Yepes (2009) ``A dynamical classification of the cosmic web'' \emph{Mon. Not. Roy. Ast.
 Soc.} \textbf{396}: 1815--1824 [ arXiv:0809.4135 ].

\bibitem{Fut88}
T Futamase (1988) 
``Approximation Scheme for Constructing a Clumpy Universe in General Relativity''
{\it Phys. Rev. Lett}. {\bf 61}, 2175-2178.


\bibitem{Fut96}
T Futamase (1996)
``Averaging of a locally inhomogeneous realistic universe''
{\it Phys. Rev}. {\bf D 53}, 681-689.

\bibitem{GasMarVen09}
M. Gasperini, G. Marozzi, G. Veneziano (2009) ``Gauge invariant averages for the cosmological backreaction ''
JCAP 0903:011[arXiv:0901.1303].

\bibitem{GasMarVen10}
M. Gasperini, G. Marozzi, G. Veneziano (2010) ``A covariant and gauge invariant formulation of the cosmological `backreaction' ''
JCAP 1002:009 [arXiv0912.3244].


\bibitem{GooLei11}
Ariel Goobar and Bruno Leibundgut (2011) ``Supernova cosmology:
legacy and future''. Invited review,  {\it Annual Review of Nuclear and
Particle Science} [arXiv:1102.1431v1].

\bibitem{goo95} J Goodman (1995). "Geocentrism reexamined", \emph{Phys Rev} \textbf{D 52},
1821 [astro-ph/9506068].

\bibitem{GosEll11}
Rituparno Goswami and George F R Ellis (2011) ``Almost Birkhoff
Theorem in General Relativity'' [arXiv:1101.4520].

\bibitem{GreWal11} Stephen R. Green and Robert M. Wald (2011)``A new framework for analyzing
the effects of small scale inhomogeneities in cosmology''
[arXiv:1011.4920].

\bibitem{Hel09} C. Hellaby (2009),
      ``Modelling Inhomogeneity in the Universe''.
{\it Proc. Sci.}  PoS (ISFTG) 005 [arXiv:0910.0350v1].

\bibitem{Hugetal01} Scott A. Hughes, Szabolcs Marka, Peter L. Bender, and Craig J.
Hogan (2001) ``New physics and astronomy with the new
gravitational-wave observatories'' eConf C010630:  P402
[arXiv:astro-ph/0110349v2].

\bibitem{Isa68a}
 Richard A. Isaacson (1968) ``Gravitational Radiation in the
Limit of High Frequency. I. The Linear Approximation and Geometrical
Optics'' \emph{Phys. Rev}. \textbf{166}, 1263--1271.

\bibitem{Isa68b}
 Richard A. Isaacson (1968) ``Gravitational Radiation in the
Limit of High Frequency. II. Nonlinear Terms and the Effective
Stress Tensor '' \emph{Phys. Rev}. \textbf{166}, 1272--1280.

\bibitem{ishetal07} M Ishak,  J Richardson, D Whittington, D., and D Garred,
(2008) "Dark Energy or Apparent Acceleration Due to a Relativistic
Cosmological Model More Complex than FLRW?" {\it Phys.Rev.} {\bf D78}:123531 [arXiv:0708.2943].

\bibitem{IshWal06}
 Akihiro Ishibashi and Robert M. Wald (2006) ``Can the
Acceleration of Our Universe Be Explained by the Effects of
Inhomogeneities?'' {\it Class.Quant.Grav.} {\bf 23}: 235-250
[arXiv:gr-qc/0509108v3].


\bibitem{Jaretal10}
N. Jarosik et (2010) ``Seven-Year Wilkinson Microwave Anisotropy Probe (WMAP]
Observations: Sky Maps, Systematic Errors, and Basic Results''  Submitted to Astrophysical Journal Supplement Series [arXiv:1001.4744].

\bibitem{JonGooMor07}
Jakob Jonsson, Ariel Goobar, Edvard Mortsell (2007) ``Tuning Gravitationally Lensed Standard Sirens'' {\it Astrophys.J}. 
{\bf 658}: 52-59 [arXiv:astro-ph/0611334]

\bibitem{Kan98} R Kantowski (1998) ``The effects of
inhomogeneities on evaluating the mass parameter $\Omega _{m}$ and
the cosmological constant $\Lambda $". \textit{Astrophys Journ}
\textbf{507}: 483-496 [ arXiv:astro-ph/9802208].


\bibitem{Kan03}
R. Kantowski (2003)
``The Lame$^{\prime}$ Equation for Distance-Redshift in Partially Filled Beam Friedmann-Lema\^{\i}tre-Robertson-Walker Cosmology''
	Phys.Rev.D68:123516 [arXiv:astro-ph/0308419v1]

\bibitem{Kan95} R Kantowski, T Vaughan and D\ Branch  (1995), ``The
effects of inhomogeneities on evaluating the deceleration
parameter". \textit{Astrophys Journ} \textbf{447}: 35-42 [arXiv:astro-ph/9511108 ].

\bibitem{KolMatRio05}
E W Kolb, S Matarrese, A Riotto (2006) ``On cosmic acceleration wihtout dark energy'': {\it New J. Phys}. {\bf 8}: 322 [arXiv:astro-ph/0506534].  

\bibitem{KolMatRio10}
Edward W. Kolb, Valerio Marra, Sabino Matarrese (2010)
``Cosmological background solutions and cosmological backreactions''
 {\it Gen.Rel.Grav.} {\bf 42}:1399-1412 [ arXiv:0901.4566]. 

\bibitem{Koretal06} A. J. Korn, F. Grundahl, O. Richard, P. S. Barklem, L.
Mashonkina, R. Collet, N. Piskunov and B. Gustafsson (2006) `A
probable stellar solution to the cosmological lithium discrepancy'
\emph{Nature} \textbf{442}, 657-659

\bibitem{GA1}  Renée C. Kraan-Korteweg and
Ofer Lahav (1998), \emph{Scientific American} (October 1998): pp.
50-57.

\bibitem{Kra97}
      A. Krasi\'nski(1997)
      {\it Inhomogeneous Cosmological Models},
      (Cambridge: Cambridge U P).

\bibitem{Kraetal10} Andrzej Krasinski, Charles Hellaby, Krzysztof Bolejko,
and Marie-Noelle C\'{e}l\'{e}rier (2010) ``Imitating accelerated expansion
of the Universe by matter inhomogeneities - corrections of some
misunderstandings'' \emph{Gen.Rel.Grav}.\textbf{42}: 2453-2475
[arXiv:0903.4070]

\bibitem{linetal95} A Linde, A., D Linde, and A Mezhlumian (1995). "Do We
Live in the Center of the World?" \emph{Phys Lett} \textbf{B 345},
203 [hep-th/9411111].

\bibitem{LinWhe57}
Richard W. Lindquist and John A. Wheeler (1957) ``Dynamics of a
Lattice Universe by the Schwarzschild-Cell Method'' \emph{Rev. Mod.
Phys}. \textbf{29}, 432–443.

\bibitem{GA2}
 C. H. Lineweaver, L. Tenorio, G. F. Smoot, P. Keegstra, A. J. Banday, and P.
Lubin (1996) ``The Dipole Observed in the COBE DMR Four-Year Data''
\emph{Astrophys.J}. \textbf{470}:38-42 [ arXiv:astro-ph/9601151].

\bibitem{LuHel07}
      T.H.-C. Lu and C. Hellaby  (2007)
      ``Obtaining the Spacetime Metric from Cosmological Observations''.
{\it Class. Quant. Grav.\/} {\bf 24}, 4107-31 [arXiv:0705.1060 ].

\bibitem{MacTau73}
 M. A. H. MacCallum and A. H. Taub (1973) ``The averaged Lagrangian and
high-frequency gravitational waves'' \emph{Comm. Math. Phys}.
\textbf{30}:153-169

\bibitem{McCHel08}
      M.L. McClure and C. Hellaby (2008),      ``Determining the Metric of the Cosmos: Stability, Accuracy, and
Consistency''.  {\it Phys. Rev. D} {\bf 78}, 044005, 1-17 [arXiv:0709.0875].

\bibitem{Mor02}
E Mortsell (2002)  
The Dyer-Roeder distance-redshift relation in inhomogeneous universes
 {\it Astron and Astrophys} {\bf 382}: 787 [astro-ph/0109197].

\bibitem{MosZibSco10} Adam Moss, James P. Zibin, and Douglas Scott (2010)
``Precision Cosmology Defeats Void Models for Acceleration''
[arXiv:1007.3725].

\bibitem{MukAbrBra97} 
V Mukhanov, L R  W Abramo, , R  Brandenberger (1997)
``On the Back Reaction Problem for Gravitational Perturbations'' {\it Phys.Rev.Lett.} {\bf 78}:1624--1627 [arXiv:gr-qc/9609026].

\bibitem{MusHelEll99} N Mustapha, C Hellaby,  and G F R Ellis  (1999).
``Large Scale Inhomogeneity vs Source Evolution: Can we Distinguish
Them?'' \emph{Mon Not Roy Ast Soc} \textbf{292}, 817
[gr-qc/9808079].

\bibitem{Noo84}
T W Noonan (1984) ``Gravitational contribution to the stress energy
tensor of a medium in GR' \emph{Gen Rel Grav} \textbf{16}:
1103-1118.

\bibitem{Peretal10}
Will J. Percival et al  (2010) ``Baryon Acoustic Oscillations in the Sloan Digital Sky Survey Data Release 7 Galaxy Sample''
{\it Mon.Not.Roy.Astron.Soc.} {\bf 401}: 2148-2168 [arXiv:0907.1660v3 ]


\bibitem{PetUza09}
P Peters and J-P Uzan (2009) \emph{Primordial Cosmology} (Oxford
University Press, USA).

\bibitem{PetBabSes11}
Antoine Petiteau, Stanislav Babak, and Alberto Sesana (2011)
``Constraining the dark energy equation of state using LISA
observations of spinning Massive Black Hole binaries''
[arXiv:1102.0769v1]

\bibitem{ReeSci67}
M J Rees, and D W Sciama (1968). "Large-scale Density
Inhomogeneities in the Universe". \emph{Nature} \textbf{217}:
511–516.

\bibitem{RegCla10} M Regis and C Clarkson (2010) "Do primordial Lithium abundances imply
there's no Dark Energy?'' [arXiv:1003.1043].

\bibitem{Rob33}
H P Robertson (1933) ``Relativistic Cosmology'' {\em Rev. Mod.
Phys.} {\bf 5}, 62.

\bibitem{San61} A R Sandage (1961) ``The Ability of the 200-Inch Telescope to
Discriminate between Selected World-Models'', \emph{Astrophys. J.}
\textbf{133}, 355.

\bibitem{SmaWil10} Peter R. Smale and David L. Wiltshire (2010) ``Supernova tests of the
timescape cosmology'' [ArXiv:1009.5855].

\bibitem{Soletal09} J. Sollerman, et al (2009) ``First-Year Sloan Digital Sky Survey-II (SDSS-II) Supernova Results: Constraints on Non-Standard Cosmological Models''  {\it Astrophysical Journal} {\bf 703 }: 1374-1385 [arXiv:0908.4276] .

\bibitem{StoEtal87}
W Stoeger, G F R Ellis, and C Hellaby (1987) ``The relationship
between continuum homogeneity and statistical homogeneity in
cosmology". \emph{Mon Not Roy Ast Soc} \textbf{226}, 373-381.

\bibitem{StoEtal07}
William R. Stoeger, Amina Helmi, and Diego F. Torres (2007)
``Averaging Einstein's Equations: The Linearized Case''
\emph{Int.J.Mod.Phys}. \textbf{D16}: 1001-1026
[arXiv:gr-qc/9904020v1].

\bibitem{StoMaaEll95}
W Stoeger, R Maartens and G F R Ellis (1995): ``Proving almost-homogeneity of the universe: 
an almost-Ehlers, Geren and Sachs theorem''. {\it Ap J} {\bf 443}, 1-5.  


\bibitem{Sze71}
Peter Szekeres (1971) ``Linearized gravitation theory in macroscopic
media'' \emph{Annals of Physics} \textbf{64}: 599-630.

\bibitem{TanNam07}
      M. Tanimoto and T. Nambu  (2007):
      ``Luminosity Distance-Redshift Relation for the LTB Solution Near
the Centre'',
      {\it Class. Quant. Grav.\/} {\bf 24}, 3843 [arXiv:gr-qc/0703012].


\bibitem{UzaClaEll08}
Jean-Philippe Uzan, Chris Clarkson, George F.R. Ellis (2008). ``Time
drift of cosmological redshifts as a test of the Copernican
principle'' \emph{Phys.Rev.Lett} \textbf{100}: 191303
[arXiv:0801.0068]

\bibitem{UzaEllLar10}
Jean-Philippe Uzan, George F.R. Ellis, and Julien Larena (2011) ``A
two-mass expanding exact space-time solution'' {\it Gen.Rel.Grav.} {\bf 43}:191-205 
 [arXiv:1005.1809].

\bibitem{vanetal06} R A Vanderveld, E A Flangan, E.A. and I Wasserman (2006).
"Mimicking Dark Energy with Lemaitre-Tolman-Bondi Models: Weak
Central Singularities and Critical Points"
  \emph{Phys. Rev.} \textbf{D 74}, 023506 [astro-ph/0602476].

 \bibitem{Wal35} A\ G\ Walker  (1935), ``On Riemannina spaces with spherical
 symmetry about a line, and the conditions for isotropy in General Relativity".
 \emph{Quart J Math Oxf} \textbf{6}, 81-93.

\bibitem{Wei72} S W Weinberg (1972), \emph{Gravitation and Cosmology} (Wiley, New
York).

\bibitem{voids_2010}
 R. van de Weygaert, K. Kreckel, E.
Platen, B. Beygu, J.H. van Gorkom, J.M. van der Hulst, M.A.
Aragon-Calvo, P.J.E. Peebles, T. Jarrett, G. Rhee, K. Kovac, and
C.-W. Yip (2010) ``The Void Galaxy Survey'' In {\it Environment and the
Formation of Galaxies: 30 years later}, Proceedings of Symposium 2
of JENAM 2010, eds. I. Ferreras, A. Pasquali, ASSP, Springer
[arXiv:1101.4187].

\bibitem{Wil07} David L. Wiltshire (2007) ``Exact solution to the averaging problem in
cosmology'' \emph{Phys.Rev.Lett}. \textbf{99}: 251101
 [arXiv:0709.0732].

\bibitem{Wil08} D L Wiltshire (2008). ``Dark energy without dark
energy'' In in \emph{Dark Matter in Astroparticle and Particle
Physics: Proceedings of the 6th International Heidelberg
Conference}, eds H.V. Klapdor-Kleingrothaus and G.F. Lewis, (World
Scientific, Singapore) pp 565-596 . [arXiv:0712.3984].

\bibitem{Wil09} David L. Wiltshire (2009) ``Average observational quantities in the timescape cosmology
'' \textit{Phys.Rev}. \textbf{D80}: 123512 [arXiv:0909.0749].

\bibitem{Wil11} David L. Wiltshire (2010) ``Gravitational energy as dark energy: cosmic structure and apparent acceleration''
to appear in the Proceedings of the Conference on Two Cosmological Models, Universidad Iberoamericana, Mexico City, 17-19 November, 2010 [arXiv:1102.2045].

\bibitem{yooetal08}
 C-M. Yoo, ,  T Kai, and   K-I Nakao (2008)
  "Solving the Inverse Problem with Inhomogeneous Universes"
 {\it Prog.Theor.Phys.} {\bf 120}:937-960 [arXiv:0807.0932].

\bibitem{Zal92} R Zalaletdinov (1992) ``Averaging Out The Einstein Equations And
Macroscopic Space-Time Geometry'' \emph{Gen.Rel.Grav}.\textbf{24}:
1015-1031.

\bibitem{Zal08}
 R Zalaletdinov (2008) ``The Averaging Problem in Cosmology and Macroscopic
 Gravity''
 \emph{Int. J. Mod. Phys}. \textbf{A 23}: 1173  [arXiv:0801.3256].

\bibitem{ZhaSte10} Pengjie Zhang and Albert Stebbins (2010) ``Confirmation of
the Copernican principle at Gpc radial scale and above from the
kinetic Sunyaev Zel'dovich effect power spectrum'' FERMILAB-PUB-10-373-A [arXiv:1009.3967].

\bibitem{StoZot92}
N V Zotov and W R Stoeger (1992)"Averaging Einstein's Equations"
\emph{Class. Quantum Grav}. \textbf{9} 1023.

\end{thebibliography}
\end{document}